\documentclass[
    reprint,
    superscriptaddress,
    aps,
    prb,
    floatfix,
]{revtex4-2}

\usepackage[utf8]{inputenc}
\usepackage[english]{babel}
\usepackage{graphicx}
\usepackage{amsmath}
\usepackage{amssymb}
\usepackage{xcolor}
\usepackage[normalem]{ulem}
\usepackage{hyperref}
\hypersetup{
    colorlinks=true,
    linkcolor=blue,
    filecolor=magenta,
    urlcolor=blue,
    citecolor=blue,
}
\usepackage{siunitx}
\usepackage[version=4]{mhchem}

\newcommand*{\tran}{^{\mkern-1.5mu\mathsf{T}}}

\begin{document}

\preprint{APS/123-QED}
\title{Limits of absolute vector magnetometry with NV centers in diamond}
\author{Dennis Lönard}
\affiliation{Department of Physics and State Research Center OPTIMAS, University of Kaiserslautern-Landau, Erwin-Schroedinger-Str. 46, 67663 Kaiserslautern, Germany}
\author{Isabel Cardoso Barbosa}
\affiliation{Department of Physics and State Research Center OPTIMAS, University of Kaiserslautern-Landau, Erwin-Schroedinger-Str. 46, 67663 Kaiserslautern, Germany}
\author{Stefan Johansson}
\affiliation{Department of Physics and State Research Center OPTIMAS, University of Kaiserslautern-Landau, Erwin-Schroedinger-Str. 46, 67663 Kaiserslautern, Germany}
\author{Jonas Gutsche}
\affiliation{Department of Physics and State Research Center OPTIMAS, University of Kaiserslautern-Landau, Erwin-Schroedinger-Str. 46, 67663 Kaiserslautern, Germany}
\author{Artur Widera}
\email{widera@physik.uni-kl.de}
\affiliation{Department of Physics and State Research Center OPTIMAS, University of Kaiserslautern-Landau, Erwin-Schroedinger-Str. 46, 67663 Kaiserslautern, Germany}
\date{\today}

\begin{abstract}
\noindent The nitrogen-vacancy (NV) center in diamond has become a widely used platform for quantum sensing. 
The four NV axes in mono-crystalline diamond specifically allow for vector magnetometry, with magnetic-field sensitivities reaching down to  \SI{}{\femto\tesla\per\sqrt{\hertz}}. 
The current literature primarily focuses on improving the precision of NV-based magnetometers. 
Here, we study the experimental accuracy of determining the magnetic field from measured spin-resonance frequencies via solving the NV Hamiltonian. 
We derive exact, analytical, and fast-to-compute formulas for calculating resonance frequencies from a known magnetic-field vector, and vice versa, formulas for calculating the magnetic-field vector from measured resonance frequencies. 
Additionally, the accuracy of often-used approximations is assessed. 
Finally, we promote using the Voigt profile as a fit model to determine the linewidth of measured resonances accurately. 
An open-source Python package accompanies our analysis. 
\end{abstract}

\maketitle

\section{Introduction}
The negatively charged nitrogen-vacancy (NV) color center in diamond has gained remarkable interest as a quantum-sensing platform \cite{barry_sensitivity_2020, aslam_quantum_2023, segawa_nanoscale_2023, scholten_widefield_2021}. 
The $S=1$ spin of the NV center leads to a splitting of the orbital ground state \ce{^3A2} under magnetic fields. 
The technique of optically detected magnetic-resonance (ODMR) spectroscopy allows the optical preparation and readout of these spin states \cite{doherty_nitrogen-vacancy_2013, gruber_scanning_1997, degen_scanning_2008}. 
Further, they can be resonantly controlled via microwave (MW) excitation and are subject to Zeeman splitting under magnetic fields \cite{doherty_negatively_2011}.
This easy access to the magnitude of the Zeeman splitting makes the NV center a promising candidate as a magnetic-field sensor \cite{rondin_magnetometry_2014}. 
Additionally, the tetrahedral structure of the diamond lattice leads to four distinct NV axes in a mono-crystalline diamond \cite{loubser_electron_1978}. 
A full ODMR spectrum, therefore, probes the magnetic-field vector along four distinct axes, enabling vector magnetometry \cite{schloss_simultaneous_2018}. 

The current literature primarily focuses on improving the readout sensitivity of NV-based magnetometers by improving fluorescence detection, diamond sample engineering or readout techniques \cite{barry_sensitivity_2020, barry_sensitive_2024} and reported sensitivities reach down to the  \SI{}{\pico\tesla\per\sqrt{\hertz}} to \SI{}{\femto\tesla\per\sqrt{\hertz}} range \cite{wolf_subpicotesla_2015, xie_hybrid_2021, wang_picotesla_2022, hu_picotesla-sensitivity_2024}. 
The reported sensitivities quantify the precision of a magnetic-field measurement, i.e., the smallest measurable relative difference of a sample field to a known magnetic field. 
However, ODMR spectra also allow for absolute magnetometry, where, instead, the accuracy of the total magnetic-field value is of interest. 
Here, we study various limitations of such absolute magnetometry and show that the accuracy is primarily determined by the uncertainties of the magnetic-field calculation. 

Calculating the magnetic field from measured ODMR resonance frequencies comprises two tasks. 
First, the magnetic-field calculation in NV coordinates for each NV axis is done.
Second, the magnetic-field angles calculated for all four axes are combined into a single total magnetic-field vector expressed in diamond lattice coordinates. 
Additionally, we also want to solve the problem of calculating the expected resonance frequencies from a given magnetic-field vector. 
For example, this can be useful in planning experiments to estimate whether a given sample exhibits a sufficiently strong magnetic field to measure it with NV-based magnetometry or to plan coil systems in an experimental setup. 

We are interested in analytical solutions to these problems that are exact, ready-to-use, easy to implement in modern programming languages, and computationally inexpensive to solve. 
Computational speed becomes essential when, for example, lock-in amplification techniques with high bandwidths are used.
DC measurement bandwidths have been demonstrated in the range of $\SI{\approx 20}{\kilo\hertz}$ \cite{schloss_simultaneous_2018} and AC measurements reach into $\SI{5}{\mega\hertz}$ bandwidths \cite{yamaguchi_bandwidth_2019, barry_sensitive_2024}. 
Calculations of the magnetic-field vector should, therefore, also be computed in time frames below $\SI{50}{\micro\second}$ for at least DC measurements. 
Lastly, we also want to estimate the accuracy of these calculations. 

In Section \ref{sec:hamiltonian}, we summarize the NV spin Hamiltonian and its characteristic polynomial used for our calculations. 
In Section \ref{sec:restob}, we first provide an analytical formula for calculating the expected spin-resonance frequencies from a given magnetic-field vector. 
Then, in Section \ref{sec:btores}, we consider the opposite problem of calculating the absolute magnetic-field value and the angle between the magnetic-field vector and the NV axis from the spin-resonance frequencies. 
In Section \ref{sec:gyro}, we highlight that the uncertainty of the NV $g$ factor is the most significant obstacle to calculating absolute magnetic fields. 
In terms of systematic errors, we discuss the validity of a widely used approximation, i.e., that the magnetic field is aligned along the NV axis, in Section \ref{sec:zeroangle}. 
Finally, in Section \ref{sec:fit}, we show that using a Voigt profile as a fit model for ODMR resonances leads to more accurate estimations of the magnetic-field sensitivity compared to Gaussian or Lorentzian fit models.

\begin{figure*}
    \centering
    \includegraphics[width=\linewidth]{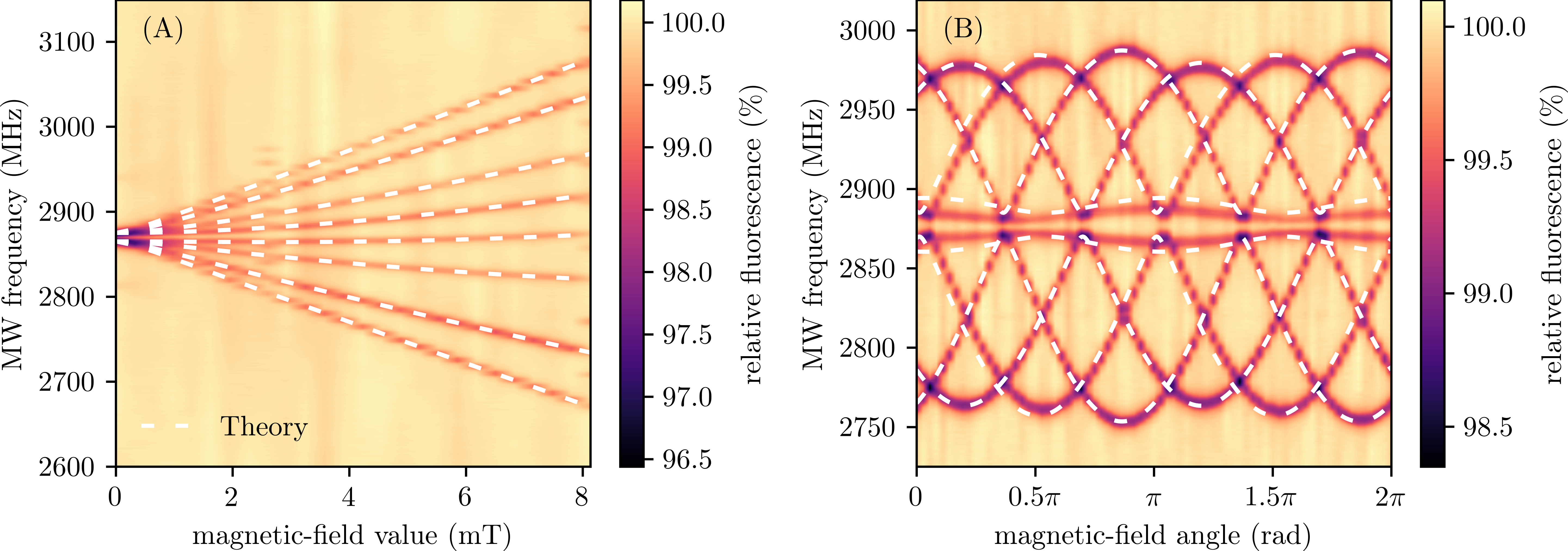}
    \caption{
        Zeeman splitting of the triplet ground states of four NV axes in a mono-crystalline diamond measured from ODMR spectra (color bar) and calculated from the analytical solution of Equation (\ref{equ:viete}) (white dashed lines). 
        The splitting is a function of (A) the total magnetic-field value and (B) the rotation of the magnetic-field vector. 
        The relations between the spin-resonance frequencies of different NV axes allow for the determination of the vector components of the magnetic field. 
        Slight deviations between the theory prediction and the measured resonances in (B) are due to alignment errors of the coil system used to generate the magnetic fields in our experiment setup. 
        Appendix \ref{sec:field_calcs} contains a detailed discussion about calculating the theory curves. 
    }
    \label{fig:splittings}
\end{figure*}

\section{Analytical solutions to the NV Hamiltonian}\label{sec:nv_solutions}
\subsection{The characteristic polynomial}\label{sec:hamiltonian}
The energy splitting of the NV ground state is accurately described by the effective Hamiltonian without hyperfine splitting \cite{doherty_theory_2012, loubser_electron_1978, manson_nitrogen-vacancy_2006}
\begin{equation}
    \label{equ:hamiltonian}
    \frac{\hat{H}}{h} = D\hat{S}_z^2 + E(\hat{S}_x^2 - \hat{S}_y^2) + \gamma_{\text{NV}}\mathbf{B} \cdot \hat{\mathbf{S}}
\end{equation}
in units of frequency, where $D \approx \SI{2870}{\mega\hertz}$ (at room temperature) denotes the zero-field splitting and $E$ is a diamond-dependent strain splitting. 
$D$ and $E$ are determined experimentally for each diamond sample and, therefore, have uncertainties that need to be included in calculations of absolute magnetometry. 
The last term describes the Zeeman splitting under magnetic fields $\mathbf{B} = B \hat{b}$, with an absolute value $B$ and a unit vector $\hat{b}$. 
To shorten the following notation, we introduce the effective magnetic-field value $\mathcal{B} = \gamma_{\text{NV}} B$ in units of frequency.
The Zeeman splitting is dependent on the gyromagnetic ratio of the NV center $\gamma_{\text{NV}}$, whose value is discussed in more detail in Section \ref{sec:gyro}. 
The $3\times3$ spin matrices are defined as 
\begin{multline}
    \hat{S}_x = \frac{1}{\sqrt{2}} \begin{pmatrix}
    0 & 1 & 0 \\
    1 & 0 & 1 \\
    0 & 1 & 0 \\
    \end{pmatrix}, \quad \hat{S}_y = \frac{1}{\sqrt{2}} \begin{pmatrix}
    0 & -i & 0 \\
    i & 0 & -i \\
    0 & i & 0 \\
    \end{pmatrix},\\
    \hat{S}_z = \begin{pmatrix}
    1 & 0 & 0 \\
    0 & 0 & 0 \\
    0 & 0 & -1 \\
    \end{pmatrix},
\end{multline}
together with the total Spin vector $\hat{\mathbf{S}} = (\hat{S}_x, \hat{S}_y, \hat{S}_z)\tran$ and the Spin number $S=1$ in the case for $\text{NV}^-$ centers \cite{loubser_electron_1978}. 

Without loss of generality, we can always rotate our coordinate system around the NV axis, so that the magnetic-field vector lies in the $xz$-plane
\begin{equation}
    \label{equ:bcoords}
    \mathbf{B} = B (\sin\theta\hat{x} + \cos\theta\hat{z}),
\end{equation}
with an angle $\theta$ between the magnetic-field vector and the NV axis. 
This eliminates all complex terms from the Hamiltonian, making it symmetric, as well as Hermitian. 
This definition is only a mathematical trick and does not require one to orient their NV sample in a specific alignment. 
The characteristic polynomial of our three level system will be a polynomial of third order with roots $\lambda_0$, $\lambda_1$ and $\lambda_2$
\begin{multline}
    \label{equ:thirdpoly}
    (\lambda - \lambda_0)(\lambda - \lambda_1)(\lambda - \lambda_2) = \\
    \lambda^3 - (\lambda_0+\lambda_1+\lambda_2)\lambda^2 + (\lambda_0\lambda_1+\lambda_1\lambda_2+\lambda_0\lambda_2)\lambda - \lambda_0\lambda_1\lambda_2.
\end{multline}
Note, that $\hat{\mathbf{S}}^2 = \hat{S}_x^2 + \hat{S}_y^2 + \hat{S}_z^2 = S(S+1)\hat{I}$, hence for a $S=1$ system $\hat{\mathbf{S}}^2 = 2\hat{I}$.
A well-known trick for such a system is to subtract a factor of $\frac{1}{3}D\hat{\mathbf{S}}^2$ from the Hamiltonian in Equation (\ref{equ:hamiltonian}) to reduce its trace to zero \cite{harari_computation_2023, balasubramanian_nanoscale_2008}. 
This trick in turn means that the characteristic polynomial reduces to the form of a depressed cubic  
\begin{equation}
    0 = \lambda^3 + p\lambda + q.
\end{equation}
This shift does not affect the resonance frequencies that we seek, because we shift all eigenvalues equally, but not relative to each other. 
The resonance frequencies are given by the roots of the characteristic polynomial
\begin{multline}
    \label{equ:char_poly}
    0 = \lambda^3 - \left(\frac{1}{3}D^2 + E^2 + \mathcal{B}^2\right)\lambda \\
    - \frac{1}{2}D\mathcal{B}^2\cos(2\theta) - E\mathcal{B}^2\sin^2\theta \\
    - \frac{1}{6}D\mathcal{B}^2 + \frac{2}{27}D^3 - \frac{2}{3}DE^2.
\end{multline} 

In the following, we use the characteristic polynomial to compute two complementary pieces of information. 
First, in Section \ref{sec:restob}, we want to find the spin-resonance frequencies for a given magnetic field by solving for the roots of the characteristic polynomial. 
Second, in Section \ref{sec:btores}, we solve for the magnetic-field vector in the case of resonance frequencies known from an ODMR spectrum. 
This is akin to knowing the roots of the characteristic polynomial and solving for its coefficients. 
We find analytical solutions for these tasks that perform significantly faster than numerically expensive optimization algorithms used frequently so far.

\subsection{Calculating the resonance frequencies from a given magnetic field}\label{sec:restob}
To find the resonance frequencies of a single NV center from a given magnetic field, we need to find the roots of the characteristic polynomial. 
In practice, the trigonometric formula due to Viète \cite{press_numerical_2007, viete_opera_1646, wolters_practical_2021, nickalls_viete_2006} is the most straightforward approach. 
We already know that Hamiltonian (\ref{equ:hamiltonian}) is symmetric, and the characteristic polynomial in Equation (\ref{equ:char_poly}) has precisely three real roots. 
Consequently, we can apply Viète's formula without checking the sign of the discriminant. 

The real roots of a depressed polynomial are then given as
\begin{equation}
    \label{equ:viete}
    \lambda_k =  \frac{2}{\sqrt{3}}\sqrt{-p}\cos\left(\frac{1}{3}\arccos\left(\frac{3\sqrt{3}}{2}\frac{q}{\sqrt{-p^3}}\right)-k\frac{2\pi}{3}\right)
\end{equation}
for $k = 0,1,2$.
Substituting 
\begin{equation}
    p = -\left(\frac{1}{3}D^2 + E^2 + \mathcal{B}^2\right)
\end{equation}
and
\begin{multline}
    q =  - \frac{1}{2}D\mathcal{B}^2\cos(2\theta) - E\mathcal{B}^2\sin^2\theta \\
    - \frac{1}{6}D\mathcal{B}^2 + \frac{2}{27}D^3 - \frac{2}{3}DE^2
\end{multline}
reveals the three roots.
The upper and lower resonance frequencies are found by $f_{\text{u}} = \lambda_2 - \lambda_0$ and $f_{\text{l}} = \lambda_1 - \lambda_0$. 
This analytical solution is exact and straightforward to evaluate. 
Figure \ref{fig:splittings} shows the resonance splitting for an example magnetic-field vector, both measured experimentally and calculated with the above formulas. 
The analytical solutions agree well with the measured resonance frequencies, except for slight deviations in the angle $\theta$, because the alignment of the magnetic field in our setup is not precisely known. 
The exact calculation is discussed in more detail in Appendix \ref{sec:field_calcs}.

\subsection{Calculating the magnetic field from given resonance frequencies}\label{sec:btores}
When the spin-resonance frequencies are known through an ODMR spectrum, for example, the magnetic field can be found analytically by comparing the characteristic polynomial of Equation (\ref{equ:char_poly}) to the roots of a general polynomial in Equation (\ref{equ:thirdpoly}) \cite{lee_vector_2015, balasubramanian_nanoscale_2008, ye_reconstruction_2019}. 
For given upper and lower resonance frequencies $f_{\text{u}}$ and $f_{\text{l}}$, the absolute magnetic-field value and the angle between the magnetic-field vector and the NV axis can be computed by
\begin{equation}
    \label{equ:b_abs}
    \mathcal{B}^2 = \frac{1}{3} \left(f_{\text{u}}^2 + f_{\text{l}}^2 - f_{\text{u}}f_{\text{l}} - D^2 - 3E^2\right)
\end{equation}
and
\begin{multline}
    \label{equ:b_theta}
    \cos^2\theta = \frac{2f_\text{l}^3-3f_\text{l}^2f_\text{u}-3f_\text{l}f_\text{u}^2+2f_\text{u}^3}{27(D-E)\mathcal{B}^2} \\
    + \frac{2D^3 - 18DE^2}{27(D-E)\mathcal{B}^2} + \frac{D-3E}{3(D-E)}.
\end{multline}
The complete derivation of these formulas can be found in Appendix \ref{sec:b_deriv}. 

For these formulas, it is assumed that the two spin-resonance frequencies of each NV axis can be uniquely identified from an ODMR spectrum. 
For specific magnetic-field vectors, it is, however, possible that spectral lines of different NV-axis overlap with each other. 
In such cases, it might be necessary to consider other fitting methods to identify $f_\text{l}$ and $f_\text{u}$, for example \cite{stone_fast_2024}, or to apply a known bias field to separate spectral lines. 

These formulas are exact for the given Hamiltonian in Equation (\ref{equ:hamiltonian}) and do not use any approximation. 
However, influences of temperature, diamond strain, electrical fields, and hyperfine splitting have been neglected in the Hamiltonian. 
In the context of magnetometry, we can assume that during a measurement, temperature, diamond strain, and outer electrical fields will be constant and therefore would not change the resulting ODMR spectrum during the measurement. 
The zero-field splitting $D$ and strain splitting $E$ constants are, however, dependent on these factors and would therefore need to be calibrated accordingly. 
Both of these constants can easily be measured with an ODMR spectrum at zero magnetic field and then substituted into Equations (\ref{equ:b_abs}) and (\ref{equ:b_theta}). 

For diamond samples where the nitrogen hyperfine splitting is resolvable, an additional term of 
\begin{equation}
    \hat{H}_\text{hfs} = \hat{\mathbf{S}}\cdot\hat{A}\cdot\hat{\mathbf{I}} + \gamma_\text{N}\mathbf{B}\cdot\hat{\mathbf{I}}
\end{equation}
would enter the Hamiltonian. 
For \ce{^15N} isotopes with $I=1/2$, this results in each spin-resonance frequency splitting up into two spectral lines. 
Otherwise, for the \ce{^14N} isotope with $I=1$, a splitting into three spectral lines will become visible. 
For magnetometry purposes, the \ce{^14N} isotope is typically used because of its higher abundance. 
In that case, the middle spectral line associated with $m_i = 0$ will not be shifted due to the hyperfine splitting and can therefore be used as the spin-resonance frequency in Equations (\ref{equ:b_abs}) and (\ref{equ:b_theta}). 
In the \ce{^15N} case, both spectral lines associated with $m_i=\pm \frac{1}{2}$ are shifted by an equal amount to a higher/lower resonance frequency. 
Here, we can take the average of both hyperfine resonance frequencies and substitute these values as $f_\text{u}$/$f_\text{l}$ into Equations (\ref{equ:b_abs}) and (\ref{equ:b_theta}). 
The additional Zeeman term that shifts the hyperfine levels is proportional to the gyromagnetic ratio of the nitrogen nucleus, which is multiple orders of magnitude smaller than the gyromagnetic ratio of the NV center ($\gamma_\text{\ce{^14N}} = \SI{3075.9(3)}{\kilo\hertz\per\tesla}$ and $\gamma_\text{\ce{^15N}} = \SI{-4315.0(4)}{\kilo\hertz\per\tesla}$ \cite{lourette_temperature_2023}). 
This nuclear Zeeman shift is even smaller than the uncertainty of the gyromagnetic ratio of the NV center, which we will discuss in Section \ref{sec:gyro}.

\subsection{Calculating the magnetic-field vector from given angles to NV axes}

\begin{figure*}
    \centering
    \includegraphics[width=\linewidth]{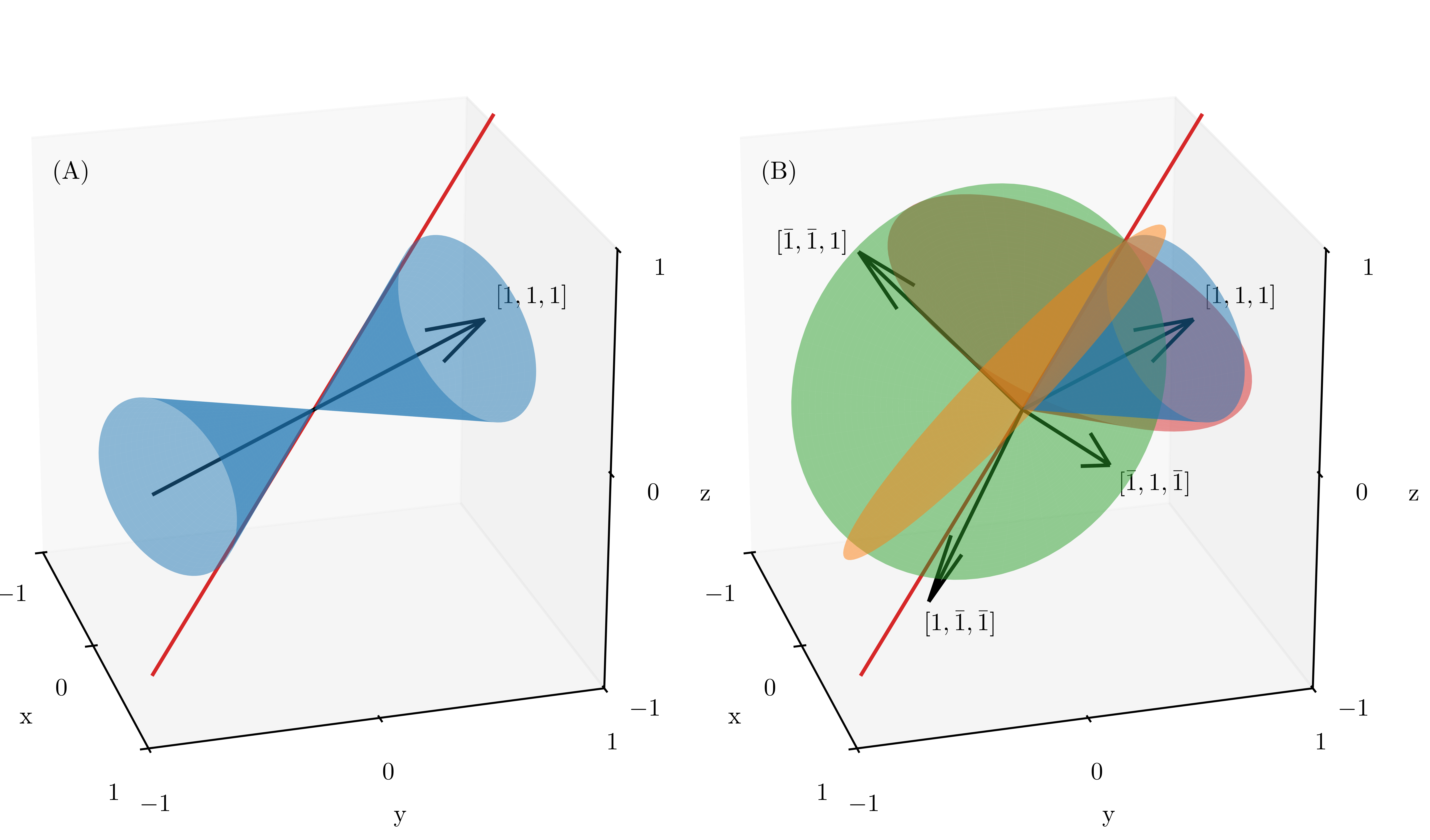}
    \caption{
        Visualization of the angles between magnetic-field vector and NV axes.
        (A) For a given magnetic-field vector (red line), the spin-resonance frequencies of the NV axis (black arrow) allow for the calculation of the angle between the two. 
        This is akin to two cones (blue surfaces) oriented along the NV axis, with the magnetic-field vector along one side of the two cones. 
        (B) When the resonance frequencies of four NV axes are known precisely, the cones (colored surfaces) will intersect along the magnetic-field vector (red line). 
        However, a real-world measurement will inevitably lead to discrepancies, and the cones will not overlap perfectly. 
        We solve this problem by computing the best linear unbiased estimator instead of the intersection in Equation (\ref{equ:bfield}). 
        The inversion symmetry of the NV axes leads to the additional problem that in the calculation, one or more cones might be oriented in the opposite direction so that no intersection exists. 
        This problem is solved by considering the sum of squared residuals in Equation (\ref{equ:ssr}). 
    }
    \label{fig:cones}
\end{figure*}

When four NV axes are probed, we find four absolute magnetic-field values and the four angles between the magnetic field and the NV axes. 
Next, we want to compute the total magnetic-field vector from our four angles $\theta_{i}$. 
Each angle yields a cylindrical cone around the corresponding NV axis, along which the magnetic-field vector can lie \cite{weggler_determination_2020}, as shown in Figure \ref{fig:cones}. 
We know that $\hat{n}_i \cdot \hat{b} = \cos\theta_i$ holds for all four angles $\theta_i$, with the NV axes $\hat{n}_i$ being parallel to the diamonds crystal axes $[111]$, $[\bar{1}\bar{1}1]$, $[\bar{1}1\bar{1}]$ and $[1\bar{1}\bar{1}]$, respectively \cite{schloss_simultaneous_2018}. 
These scalar products reveal a linear set of four equations $N \cdot \hat{b} = c$ with three unknowns, i.e., the three vector components of the magnetic-field unit vector $\hat{b}$
\begin{equation}
    \frac{1}{\sqrt{3}} \begin{pmatrix}
    1 & 1 & 1 \\
    -1 & -1 & 1 \\
    -1 & 1 & -1 \\
    1 & -1 & -1\\
    \end{pmatrix} \cdot \hat{b} = \begin{pmatrix}
    \pm\cos\theta_1 \\
    \pm\cos\theta_2 \\
    \pm\cos\theta_3 \\
    \pm\cos\theta_4 \\
    \end{pmatrix}.
\end{equation}
This over-determined set of linear equations is akin to a fit problem, with a model function that is linear in its parameters. 
The best linear unbiased estimator (BLUE) minimizes the sum of squared residuals (SSR) and is expressed as \cite{press_numerical_2007} 
\begin{equation}
    \hat{b}_{\text{BLUE}} = (N^TN)^{-1}N^T c.
\end{equation}
A single matrix multiplication now reconstructs the total magnetic-field vector
\begin{equation}
    \label{equ:bfield}
    \mathbf{B} = B \frac{\sqrt{3}}{4} \begin{pmatrix}
    1 & -1 & -1 & 1 \\
    1 & -1 & 1 & -1 \\
    1 & 1 & -1 & -1 \\
    \end{pmatrix} \cdot \begin{pmatrix}
    \pm\cos\theta_1 \\
    \pm\cos\theta_2 \\
    \pm\cos\theta_3 \\
    \pm\cos\theta_4 \\
    \end{pmatrix}.
\end{equation}
If we additionally use Gaussian error propagation to calculate the variances $\sigma^2_{i}$ of each $\cos\theta_i$, we can instead use the generalization of weighted sums due to Aitken \cite{aitken_least_1936}
\begin{equation}
    \label{equ:aitken}
    \hat{b}_{\text{BLUE}} = (N^TWN)^{-1}N^TWc,
\end{equation}
with a diagonal weight matrix $W_\text{ii} = 1 / \sigma^2_{i}$. 

Because of the ambiguities in $\pm\cos\theta_i$, we have $2^4$ possible solutions $\hat{b}$. 
For perfect measurements, at least two combinations of cones exist with precisely one intersection. 
Slight measurement errors will lead to the problem of the cones not intersecting perfectly. 
This problem is often avoided in literature by measuring only three NV axes. 
The solution of Equation (\ref{equ:bfield}) solves this problem analytically by not computing the intersection but the vector that most closely fits a supposed intersection. 
Nonetheless, we still have $2^4$ possible solutions. 
However, most of these solutions represent four cones that do not intersect because one or more cones are oriented incorrectly. 
Unwanted solutions will have a large SSR
\begin{equation}
    \label{equ:ssr}
    S(\hat{b}) = (c-N\hat{b})^T (c-N\hat{b}).
\end{equation}
To find the correct solution, we compute all permutations of Equation (\ref{equ:bfield}) and sort them by their SSR. 
The vector $\hat{b}$ that minimizes $S(\hat{b})$ is then our final solution. 

Finally, we do not know which spin-resonance frequency in the ODMR spectra belongs to which NV axis, meaning we only know the magnetic-field vector up to the diamond lattice's $\text{T}_{\text{d}}$ symmetry. 
Additionally, we do not know the direction of the Zeeman splitting, adding an additional inversion symmetry i. 
This symmetry is an inherent property of all NV-based magnetometers. 
To obtain all possible magnetic-field vectors that reproduce identical ODMR spectra, we must multiply our solution with all 48 transformation matrices of the $\text{O}_\text{h}$ symmetry group listed in Appendix \ref{sec:symmetry}. 
However, this problem can also be circumvented by applying a known bias field, which would break the symmetry and thus reduce the possible magnetic-field vectors to one unambiguous solution. 

To summarize, the recipe for calculating the magnetic field from given spin-resonance frequencies goes as follows. 
First, Equation (\ref{equ:b_abs}) and (\ref{equ:b_theta}) are used to calculate the absolute magnetic-field strength $\mathcal{B}$ and the angle of the magnetic-field vector to the NV axes $\theta$. 
These equations include the strain splitting $E$ and are thus more generally applicable to any NV-diamond sample than previous approaches reported \cite{beaver_optimizing_2024, silani_nuclear_2023, weggler_determination_2020, ye_reconstruction_2019}. 
Furthermore, $\mathcal{B}$ and $\theta$ are the variables of interest, compared to polar and azimuth angles computed by other similar formulas \cite{balasubramanian_nanoscale_2008}. 
The crucial difference between our derivation and the approximations assuming $D \gg E$, as used, for example, in \cite{balasubramanian_nanoscale_2008}, is that we have taken the symmetry of the NV center into account in Equation (\ref{equ:bcoords}). 
This symmetry then led us to a solvable system of only two variables, i.e., the magnetic-field value $\mathcal{B}$ and the angle to the NV axis $\theta$, instead of an unsolvable system with three variables that has to be approximated. 

Second, with Equation (\ref{equ:bfield}) and (\ref{equ:ssr}), the magnetic-field vector is calculated in diamond-lattice coordinates without approximations \cite{simin_high-precision_2015, maertz_vector_2010}, the need to discard one NV axis, numerically expensive fitting \cite{steinert_high_2010, garsi_three-dimensional_2024} or machine-learning algorithms \cite{tsukamoto_accurate_2022, homrighausen_edge-machine-learning-assisted_2023, zhang_deep-neural-network-based_2024}. 
Transforming from diamond-lattice coordinates to laboratory coordinates can be done with a simple rotation matrix but is dependent on the orientation of the diamond in any given experiment. 
Lastly, we discussed the symmetry of the computed magnetic-field vector. 
Whether the calculation of additional magnetic-field vectors is necessary depends on whether one is interested in only one or all solutions, depending on the application.

\section{Accuracy considerations}
While the above approach allows to compute magnetic-field properties from measured spin-resonance frequencies or vice versa, these values are prone to systematic shifts, including knowledge of the gyromagnetic ratio, the alignment of the magnetic field relative to the diamond crystal, or the precise form of the fitting function to determine the resonance frequencies. 

Measurement sensitivities are broadly divided into two categories. 
The relative magnetic-field sensitivity describes the smallest change in magnetic-field value that a sensor can measure, given in units of \SI{}{\tesla\per\sqrt\hertz}. 
It is limited only by correlations of the statistical noise that depends on each sensor and is therefore used as a benchmark in literature discussing the design and implementation of magnetic-field sensors. 

In this Section, we want to discuss the absolute accuracy of a magnetic-field measurement. 
It describes instead how accurately the readout of the sensor corresponds to the actual magnetic-field value, given in units of \SI{}{\tesla}. 
The absolute accuracy is only limited by systematic uncertainties of how the magnetic-field value is computed from given spin-resonance frequencies. 
The systematic uncertainties are sensor-independent and cannot be averaged out over multiple measurements. 

In the following, we will estimate that the limiting factor of such systematic uncertainties is the uncertainty of the value of the gyromagnetic ratio $\gamma_\text{NV}$. 
The uncertainty in the order of $10^{-4}$ of the gyromagnetic ratio directly results in an uncertainty of $10^{-4}$ when calculating a magnetic-field value. 
For example, when a magnetic-field value of \SI{10}{\milli\tesla} is measured, the actual magnetic-field value can only be determined up to an uncertainty of roughly \SI{}{\micro\tesla}.

\subsection{Uncertainty of the gyromagnetic ratio}\label{sec:gyro}
Any calculation of the magnetic field is always linearly dependent on the gyromagnetic ratio of the NV center 
\begin{equation}
    \gamma_{\text{NV}} = \frac{g_{\text{NV}}\mu_{\text{B}}}{h} = \SI{28.032(4)}{\giga\hertz\per\tesla}.
\end{equation}
The Bohr magneton $\mu_{\text{B}}$ is given in the literature to such high degrees of precision that we do not need to take its uncertainty into account for our purposes, and the Planck constant $h$ is an exact quantity \cite{tiesinga_codata_2021}. 
However, the effective $g_{\text{NV}}$-factor of the NV center differs from the gyromagnetic ratio of the free electron $g_e$ due to the Coulomb potential of the nitrogen atom \cite{doherty_theory_2012, doherty_negatively_2011}. 
Therefore, $g_{\text{NV}}$ has to be measured experimentally and has a significant associated uncertainty. 
Furthermore, the $C_{3v}$ symmetry class of the NV center dictates that an anisotropy can exist between the Zeeman components of magnetic fields parallel $g_\parallel$ and perpendicular $g_\perp$ to the NV axis. 
This can be described by an effective $\bar{g}_{\text{NV}}$-matrix 
\begin{equation}
    \bar{g}_{\text{NV}} = 
    \begin{pmatrix}
        g_\perp & 0 & 0 \\
        0 & g_\perp & 0 \\
        0 & 0 & g_\parallel \\
    \end{pmatrix}.
\end{equation}
The absolute value of the $g_{\text{NV}}$-factor is well established by experiments, but the anisotropy is so slight that it is barely measurable, as shown in Table \ref{tab:gfactor}. 

\begin{table}
    \centering
    \caption{
        Experimentally measured $g_{\text{NV}}$-factor, adapted from \cite{doherty_theory_2012}. 
        The associated uncertainties need to be considered to achieve accurate magnetic-field values.
        However, we estimate the anisotropy between $g_\perp$ and $g_\parallel$ to be less impactful than the overall uncertainty. 
    }
    \begin{tabular}{lcc}
        \hline\hline
        Reference & $g_\perp$ & $g_\parallel$ \\
        \hline
        Loubser \& Wyk \cite{loubser_electron_1978} & \SI{2.0028\pm0.0003}{} & \SI{2.0028\pm0.0003}{} \\
        He \cite{he_paramagnetic_1993} & \SI{2.0028\pm0.0003}{} & \SI{2.0028\pm0.0003}{} \\
        Felton \cite{felton_hyperfine_2009} & \SI{2.0031\pm0.0002}{} & \SI{2.0029\pm0.0002}{} \\
        \hline\hline
    \end{tabular}
    \label{tab:gfactor}
\end{table}

We want to estimate the influence of the uncertainty and the anisotropy of the $g_{\text{NV}}$-factor on an absolute magnetic-field measurement of a \SI{10}{\milli\tesla} field. 
We use this specific magnetic-field value because it sufficiently separates the spin-resonance frequencies to make them distinguishable while avoiding the need for excessively low or high microwave frequencies to record an ODMR spectrum.
Moreover, most NV magnetometry takes place in roughly this order of magnitude of magnetic-field strengths. 
When we assume no anisotropy and a value of $g_{\text{NV}} = \SI{2.0028\pm0.0003}{}$, the uncertainty of a magnetic-field measurement of a \SI{10}{\milli\tesla} field is \SI{\approx 1.5}{\micro\tesla}. 
To estimate the influence of the anisotropy, we numerically solve the Hamiltonian for a \SI{10}{\milli\tesla} field at an angle of \SI{1}{\degree} to the NV axis with the numerical method described in Appendix \ref{sec:benchmarks}. 
Using $g_\perp = g_\parallel = \SI{2.0030}{}$ versus $g_\perp = \SI{2.0031}{}$ and $g_\parallel = \SI{2.0029}{}$ we find a discrepancy of \SI{\approx 0.5}{\micro\tesla}. 
We conclude that the limiting factor of absolute vector magnetometry with NV centers is the uncertainty of the absolute value of the $g_{\text{NV}}$-factor rather than its anisotropy. 
Therefore, we do not consider the anisotropy when deriving our analytical solutions. 

To combat the uncertainty of the gyromagnetic ratio, various setups for referencing can be employed. 
For example, the to-be-measured magnetic field can be compared to a known bias field, or cross-checks against other magnetometry protocols, like NMR or fluxgates, can be performed. 

\subsection{Relative alignment between magnetic-field vector and NV axes}\label{sec:zeroangle}
The magnetic field is intentionally aligned along the NV axis for many applications, i.e., $\theta \approx 0$. 
For example, in single NV experiments, the magnetic field can be aligned by maximising the observed ODMR splitting. 
Regarding NV ensembles, many experiments choose one NV axis and align their magnetic field by overlapping the ODMR resonance lines of the other three NV axes. 
No matter how the magnetic field is aligned, there will always be some slight misalignment angle $\theta$ left. 
In this context, of assumed alignment, the Zeeman splitting is often approximated to be linearly dependent on the magnetic field \cite{homrighausen_microscale_2024, beaver_optimizing_2024, silani_nuclear_2023, scholten_widefield_2021}, and the magnetic-field value is calculated as
\begin{equation}
    \label{equ:b_approx}
    \mathcal{B} = \sqrt{\frac{(f_{\text{u}}-f_{\text{l}})^2}{4} - E^2}.
\end{equation}
When assuming a linear splitting, this formula is derived from Equation (\ref{equ:b_abs}). 

\begin{figure}
    \centering
    \includegraphics[width=\linewidth]{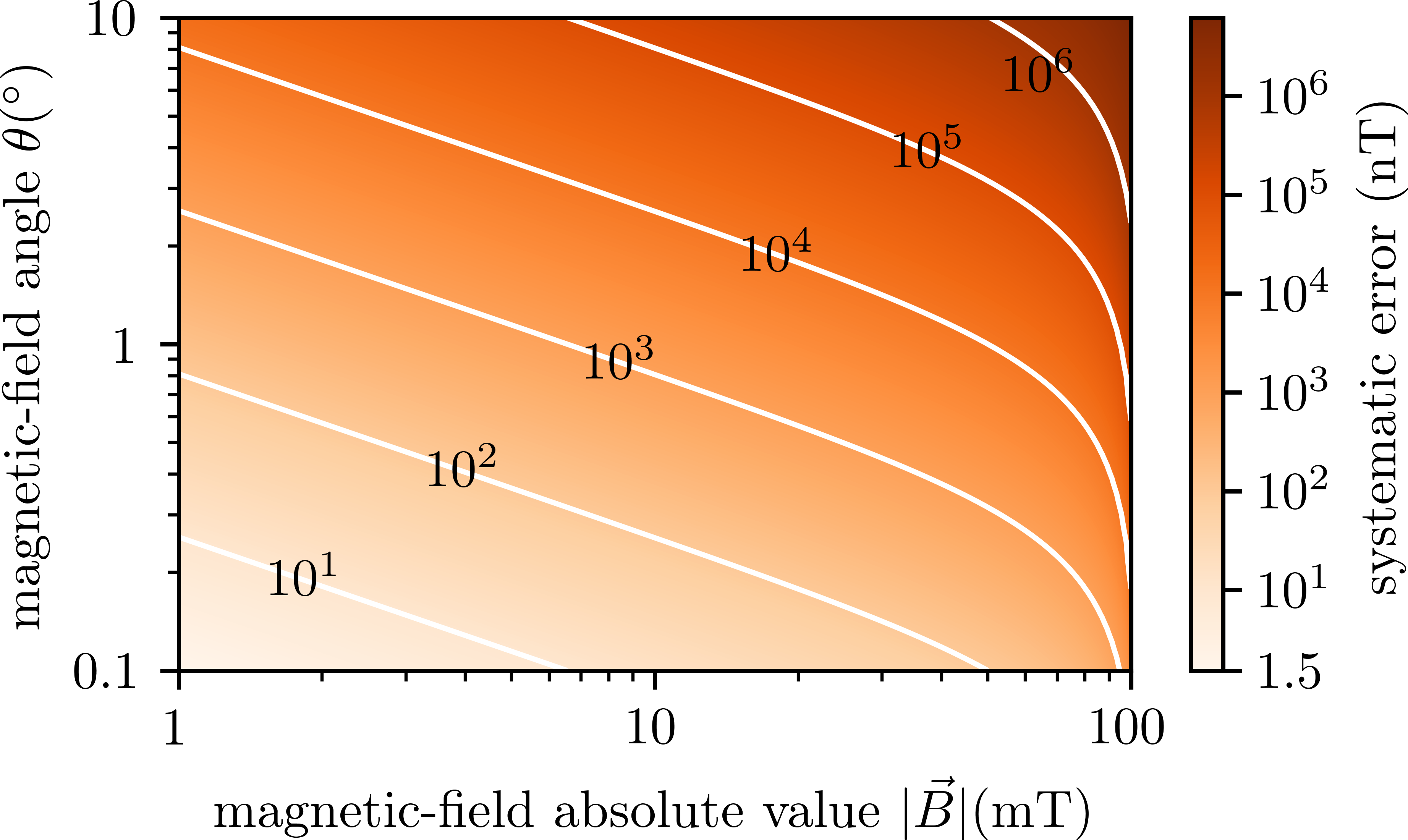}
    \caption{
        Relative systematic error of the approximation that the magnetic field is aligned with the NV axis in Equation (\ref{equ:b_approx}) when the NV axis is, in reality, slightly misaligned by an angle $\theta$. 
        Even for small magnetic fields and small angles $\theta$, the relative error of the assumption is far more significant than the typical sensitivities of NV-based sensors.
        For example, when measuring an absolute magnetic field of $\SI{10}{\milli\tesla}$, at an angle of $\SI{1}{\degree}$ to the NV axis, the relative error is $\SI{\approx 1.5}{\micro\tesla}$ and therefore larger than typical magnetic-field sensitivities. 
    }
    \label{fig:theta_error}
\end{figure}

Regarding accuracy, the question arises of how well the magnetic field is aligned along the NV axis. 
If we assume the alignment to be perfect, but in reality there is still a misalignment angle of let's say \SI{1}{\degree} present, how will this affect the calculated magnetic-field value? 
We calculate the relative systematic error between the exact formula in Equation (\ref{equ:b_abs}) and the approximation in Equation (\ref{equ:b_approx}) as a function of the misalignment angle $\theta$ and the magnetic-field absolute value and plot it in Figure \ref{fig:theta_error}. 

Even for applications where the magnetic-field vector is considered to be aligned with the NV axis, technical limitations often lead to slight misalignments in the order of \SI{\approx 1}{\degree}. 
In this case, when a magnetic-field value of \SI{10}{\milli\tesla} is measured, the above assumption would lead to a systematic error of \SI{\approx 1.5}{\micro\tesla}. 
When accuracies of $< \SI{}{\micro\tesla}$ are required, this result suggests using the exact formula of Equation (\ref{equ:b_abs}), even when the magnetic field is considered to be aligned with the NV axis because even slight deviations in the angle $\theta$ will lead to significant systematic errors that can be orders of magnitude larger than the magnetic-field sensitivity. 

\subsection{Comparison of different uncertainties}
The splitting constants $D$ and $E$ and the spin-resonance frequencies $f_\text{u}$ and $f_\text{l}$ are typically determined experimentally, usually with fit curves to ODMR spectra. 
The uncertainties of these quantities are then given by the inverse of the covariance matrix of the fitting algorithm and add to the uncertainty of the $g_\text{NV}$-factor and possibly misaligned magnetic-field vectors. 
These uncertainties are then dependent on many factors, like the specific noise in the spectra, making it difficult to estimate the exact range of uncertainties. 

Additionally, $D$, $E$, $f_\text{u}$, and $f_\text{l}$ are also dependent on environmental factors like temperature, strain, and electric fields. 
However, on the one hand, these factors do not influence the spin-resonance frequencies to the same degree as magnetic fields do (\SI{-74.2(7)}{\kilo\hertz\per\kelvin} \cite{acosta_temperature_2010}, \SI{14.58(6)}{\mega\hertz\per\giga\pascal} \cite{doherty_electronic_2014}, and \SI{17(3)}{\hertz\centi\meter\per\volt} \cite{dolde_electric-field_2011}). 
On the other hand, we are focusing here on high bandwidth sensors that can measure magnetic-field values in time frames of \SI{\approx 50}{\micro\second}, so these environmental factors would have to change significantly during these time frames to be regarded as a significant uncertainty. 
In cases where this is a concern, measurement protocols exist, like multiplexed sensing \cite{shim_multiplexed_2022}, that can distinguish between a magnetic-field change and any other influence. 
The recipe for calculating the magnetic-field vector presented in Section \ref{sec:nv_solutions} still applies to these measurement protocols. 

We roughly estimate, from multiple fit curves, that the deviations of the fitted constants $D$, $E$, $f_\text{u}$ and $f_\text{l}$ are typically at an order of magnitude of \SI{\approx 0.01}{\mega\hertz}. 
Assuming a \SI{10}{\milli\tesla} magnetic-field amplitude was measured, we use Gaussian error propagation in Equation (\ref{equ:b_abs}) to estimate the resulting magnetic-field-value uncertainty of each of these values in Table \ref{tab:uncertainties}. 
Each fit uncertainty leads to an uncertainty in the order of \SI{}{\micro\tesla}, except for the uncertainty of $E$, because $E \ll D$. 
However, this holds only for isotopically impure diamond samples, in which the hyperfine splitting is not visible. 
If, instead, isotopically pure samples are used, the linewidth of ODMR resonances is typically one order of magnitude smaller. 
Therefore, the fit uncertainties of $D$, $E$, $f_\text{u}$ and $f_\text{l}$ will also be one order of magnitude smaller than what we show here in Table \ref{tab:uncertainties}. 
We conclude that while relative magnetometry with NV-based sensors is limited by the magnetic-field sensitivity of each sensor in typical ranges of \SI{}{\pico\tesla\per\sqrt{\hertz}} to \SI{}{\femto\tesla\per\sqrt{\hertz}}, absolute magnetometry will mainly be limited by the uncertainty of the $g_\text{NV}$-factor independent of the sensor. 

\begin{table}
    \centering
    \caption{
        Estimated influence on the calculated magnetic-field value of different uncertainties. 
        Assuming a magnetic-field value of \SI{10}{\milli\tesla} is measured, the uncertainty of a calculated magnetic-field value lies in the order of \SI{}{\micro\tesla}. 
        The most significant contributions stem from the uncertainty of the $g_{\text{NV}}$-factor, a systematic error of approximating $\theta \approx \SI{0}{\degree}$, when the field might be misaligned, and from the uncertainty of the $D$ splitting parameter. 
    }
    \begin{tabular}{cc}
        \hline\hline
        Measurement & Relative magnetic-field  \\
        uncertainty & uncertainty \\
        \hline
        $g_{\text{NV}}$ uncertainty & \SI{\approx 15e-3}{} \\  
        $g_{\text{NV}}$ anisotropy & \SI{\approx 5e-3}{} \\  
        $\theta \approx \SI{0}{\degree}$ approximation & \SI{\approx 15e-3}{} \\  
        $D$ fit uncertainty & \SI{\approx 12e-3}{} \\  
        $E$ fit uncertainty & \SI{\approx 60e-6}{} \\  
        $f_\text{u}$ / $f_\text{l}$ fit uncertainty & \SI{\approx 8e-3}{} \\  
        \hline\hline
    \end{tabular}
    \label{tab:uncertainties}
\end{table}

\section{Computation time}\label{sec:comptime}
\begin{figure}
    \centering
    \includegraphics[width=\linewidth]{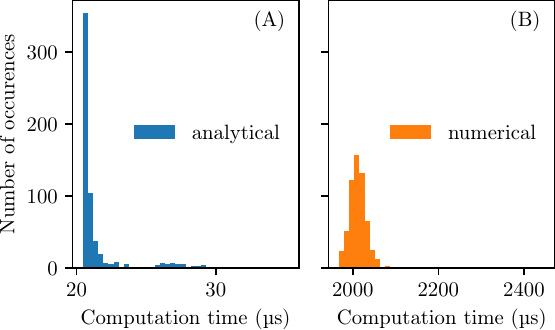}
    \caption{
        Benchmarks for computation time of (A) analytical approach presented here and (B) a typical numerical approach. 
        DC readout bandwidths of NV magnetometers are in the range of $\SI{\approx 20}{\kilo\hertz}$, requiring computation times of the magnetic-field vector in the order of $\SI{50}{\micro\second}$. 
        Our analytical approach outperforms a typical numerical approach and can determine the magnetic-field vector faster than the resonances are usually measured. 
    }
    \label{fig:comptime}
\end{figure}

For DC fields, NV-based magnetometers can reach measurement bandwidths in the range of up to $\SI{\approx 20}{\kilo\hertz}$ \cite{schloss_simultaneous_2018}. 
When the spin-resonance frequencies are measured in time frames of roughly $\SI{50}{\micro\second}$, we must compute the magnetic-field vector faster than the readout rate to avoid limiting the measurement bandwidth. 
In Figure \ref{fig:comptime}, we compare the computation time of our analytical approach presented here to a typical numerical approach of minimizing a cost function that depends on an initially guessed magnetic-field vector \cite{homrighausen_microscale_2024}. 
We see that the analytical approach reaches computation times of $\SI{\approx 20}{\micro\second}$, fast enough for typical NV sensor bandwidths, and outperforming the numerical approach, which takes in the order of $\SI{\approx 2000}{\micro\second}$ for the same task. 
Additionally, the analytical solution does not require any guesses of initial parameters compared to the numerical approach. 
Appendix \ref{sec:benchmarks} contains more details about how the benchmarks were performed. 

\section{Fit model for ODMR Spectra}\label{sec:fit}
A standard procedure to characterise the relative magnetic-field sensitivity in NV magnetometry literature is to derive a shot-noise-limited sensitivity $\eta$ from a measured ODMR spectrum. 
In this section, we discuss the influence of the fit model used on the calculated shot-noise-limited sensitivity. 
For this, and many other applications, the resonance frequency $f_{\text{res}}$, the contrast $C$, and the full-width half-maximum (FWHM) linewidth $\alpha$ are extracted from fit models to normalized ODMR spectra that scan over the MW frequency $f$. 
For weak MW driving, as we average over an ensemble of many NV centers, the spectral line will have a Gaussian shape \cite{dreau_avoiding_2011}
\begin{equation}
    G(f, C_G, \alpha_G) = 1 - C_G\exp\left(4\ln\left(\frac{1}{2}\right)\frac{(f-f_\text{res})^2}{\alpha_G^2}\right).
\end{equation}
However, as the MW driving strength increases, power broadening becomes relevant, which distorts the line shape towards a Lorentzian of the form \cite{citron_experimental_1977, vitanov_power_2001}
\begin{equation}
    L(f, C_L, \alpha_L) = 1 - \frac{C_L\alpha_L^2}{4(f-f_{\text{res}})^2 + \alpha_L^2}.
\end{equation}
The exact lineshape is then dependent on several experimental factors, including whether single NV centers or ensembles are used, the laser power, the microwave power, or the size of the diamond sample. 

Typically, the best magnetic-field sensitivities are reached when the contrast is maximal and the linewidth begins to broaden.
Consequently, neither a pure Gaussian nor a pure Lorentzian line shape can be expected to fit the data. 
Lorentzian fits then underestimate the FWHM linewidth, while Gaussian fits overestimate the linewidth. 
Instead, we study the so-called Voigt profile $V(x)$, defined as the convolution of a Lorentzian $L(x)$ and a Gaussian $G(x)$
\begin{equation}
    V(x, C_V, \alpha_L, \alpha_G) = \int_{\mathbb{R}} G(\tau, C_G, \alpha_G)L(x-\tau, C_L, \alpha_L) \,d\tau.
\end{equation}

\begin{figure}
    \centering
    \includegraphics[width=\linewidth]{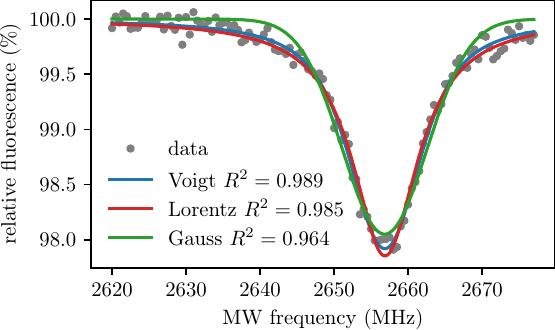}
    \caption{
        An example of an ODMR resonance fitted with a Gaussian, a Lorentzian, and a Voigt profile.
        The Voigt profile achieves the highest $R^2$ value. 
        Calculating ODMR sensitivities from these fit curves leads to significant differences, i.e., $\eta_\text{Voigt} = \SI{2.65}{\micro\tesla\per\sqrt{\hertz}}$, $\eta_\text{Lorentz} = \SI{2.45\pm0.05}{\micro\tesla\per\sqrt{\hertz}}$ and $\eta_\text{Gauss} = \SI{3.05\pm0.08}{\micro\tesla\per\sqrt{\hertz}}$. 
        The Lorentz fit in this example overestimates the actual sensitivity by more than $\SI{8}{\percent}$. 
    }
    \label{fig:voigt}
\end{figure}

\begin{figure*}
    \centering
    \includegraphics[width=\linewidth]{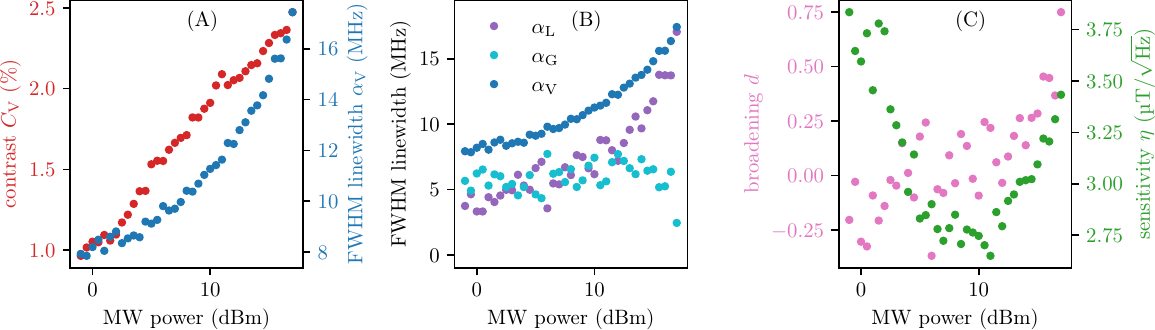}
    \caption{
        Voigt fit parameters describing the power broadening of the example ODMR peak shown in Figure \ref{fig:voigt}. 
        (A) Both contrast and FWHM linewidth increase with higher MW power, but it is unclear which MW power would lead to an optimal magnetic-field sensitivity. 
        (B) The individual linewidths of the Lorentzian and the Gaussian part of the Voigt fit indicate how far the power broadening increases the total linewidth. 
        For higher MW powers, the line shape is shifted to a Lorentzian shape, where $\alpha_\text{L} \rightarrow \alpha_\text{V}$ and $\alpha_\text{G} \rightarrow 0$. 
        (C) The broadening factor $d$ can describe the power broadening, showing the Lorentzian shape again with $d \rightarrow 1$. 
        The magnetic-field sensitivity $\eta$, dependent on contrast and linewidth $(\propto \alpha_\text{V} / C_\text{V})$, reaches a minimum when the power broadening starts to take effect. 
        In this case, the magnetic-field sensitivity is only in the order of \SI{}{\micro\tesla} because a diamond with a low nitrogen concentration was used for these ODMR spectra. 
        The nitrogen concentration does not affect the line broadening. 
    }
    \label{fig:voigtfit}
\end{figure*}

The magnetic-field sensitivity is a function of the maximum value of the derivative of the ODMR signal and the photon-collection rate $P$ \cite{barry_sensitive_2024}
\begin{equation}
    \eta = \frac{1}{\gamma_\text{NV}} \frac{1}{\text{max}\left|\frac{\partial \text{ODMR}(f)}{\partial f}\right|} \frac{1}{\sqrt{P}}
\end{equation}
and is therefore dependent on which fit model is used to fit the data. 
Figure \ref{fig:voigt} shows the ODMR spectrum with optimal MW power, i.e., minimal sensitivity, from the data set shown in Figure \ref{fig:voigtfit}. 
The Voigt profile achieves the highest $R^2$ value, indicating that it models the line shape more accurately than the Lorentzian and the Gaussian profiles. 
When magnetic-field sensitivities are derived from the contrast and the linewidth of these fit profiles, the Lorentz profile significantly overestimates the achieved sensitivity, in this example, by more than \SI{8}{\percent}. 

The Voigt profile can be expressed by the real part of the Faddeeva function \cite{shippony_highly_1993, shippony_correction_2003, weideman_computation_1994}
\begin{equation}
    \omega(z) = e^{-z^2}\text{erfc}(-iz),
\end{equation}
which is readily available in commonly used programming languages like Python \cite{the_scipy_community_scipy_2024}, C++ \cite{johnson_c_2025}, and MATLAB \cite{johnson_matlab_2025}. 
The fit model is then written as 
\begin{equation}
    V(f) = 1 - \frac{\text{Re}[\omega(z)]}{\sigma\sqrt{2\pi}}, \enspace\text{with}\enspace z = \frac{f - f_{\text{res}} + i\nu}{\sigma\sqrt{2}},
\end{equation}
with $\nu$ and $\sigma$ as width parameters for the Lorentzian and Gaussian parts, respectively.
In lock-in amplified measurements using frequency modulation, the derivative of the ODMR signal is recorded instead, and the fit model becomes 
\begin{equation}
    \frac{\partial V(f)}{\partial f} = -\frac{1}{\sigma^3\sqrt{2\pi}}\left(\nu\text{Im}[\omega(z)] - (f - f_{\text{res}})\text{Re}[\omega(z)]\right).
\end{equation}

The FWHM linewidths of the Lorentzian and the Gaussian part of the model are retrieved by
\begin{align}
    \alpha_\text{G} &= 2\sigma\sqrt{2\ln 2}, \\
    \alpha_\text{L} &= 2\nu.
\end{align}
The only inconvenience of using the Voigt profile as a fit model is that the total FWHM linewidth $\alpha_\text{V}$ is not trivial to calculate.
However, simple-to-use numerical approximations \cite{olivero_empirical_1977, whiting_empirical_1968, kielkopf_new_1973} like
\begin{equation}
    \alpha_\text{V} \approx 0.5346 \alpha_\text{L} + \sqrt{0.2166 \alpha_\text{L}^2 + \alpha_\text{G}^2}.
\end{equation}
exist, that are sufficiently accurate (relative error $\SI{<0.025}{\percent}$). 
With the Voigt profile, we can describe the line broadening due to MW power saturation with the dimensionless coordinate \cite{olivero_empirical_1977} 
\begin{equation}
    d = \frac{\alpha_\text{L}-\alpha_\text{G}}{\alpha_\text{L}+\alpha_\text{G}} \enspace\in[-1, 1].
\end{equation}

Figure \ref{fig:voigtfit} shows the Voigt fitting parameters for an example ODMR resonance with varying MW power. 
As expected, the contrast of the resonance increases in Figure \ref{fig:voigtfit} (A), but so does the linewidth. 
It is not immediately apparent which MW power will lead to an optimal sensitivity. 
However, the Voigt fit parameters $\sigma$ and $\nu$ give a good indication of the line broadening in Figure \ref{fig:voigtfit} (B), where for large MW powers the Lorentzian part of the linewidth $\alpha_\text{L}$ converges to the total linewidth $\alpha_\text{V}$, while the Gaussian part of the linewidth $\alpha_\text{G}$ decreases towards zero. 
When we compare the broadening factor $d$ to the sensitivity $\eta$ in Figure \ref{fig:voigtfit} (C), we can see that the sensitivity worsens significantly when the broadening factor increases or decreases drastically. 
A typical and time-consuming procedure to optimize the sensitivity is to sweep the MW power and record multiple ODMR spectra, apply fit curves, calculate the sensitivity, and see which MW power is optimal. 
A clear benefit of using a Voigt profile is that a single ODMR spectrum is sufficient to calculate the broadening factor $d$ and, therefore, see whether the applied MW power was too low or too high. 
With this additional information, one can quickly narrow down the right region of MW power for a given laser power.

\section{Conclusion}
Here, we provided ready-to-use formulas for accurately and analytically solving the NV Hamiltonian for calculating magnetic fields from measured spin-resonance frequencies and predicting the resonance splitting from a given magnetic field. 
We showed that these formulas are computationally inexpensive to solve, which can drastically improve the readout bandwidth of NV-based magnetometers. 

We discussed the uncertainty and limitations of absolute vector magnetometry with NV centers in diamonds by estimating the influence of different measurement uncertainties. 
We concluded that the uncertainty of absolute magnetometry typically lies in the order of a few \SI{}{\micro\tesla}. 
The most significant limitation is the uncertainty of the $g_{\text{NV}}$-factor. 
Additionally, the assumption that the magnetic-field vector is perfectly aligned with the NV axis leads to a significant systematic error, when the magnetic field is even slightly misaligned. 

Finally, we discussed the influence of the fit model used for ODMR spectra on the calculated shot-noise-limited sensitivity. 
We showed that the Voigt profile more accurately fits recorded ODMR spectra than Gaussian or Lorentzian fit models and brings the additional benefit of detecting the MW power broadening from a single ODMR spectrum.

\section{Data and code availability}
All data shown in the figures are available on Zenodo \cite{lonard_data_2025}. 
The authors published a Python package containing routines to calculate the analytical solutions presented on the Python Package Index (PyPI) \cite{lonard_python_2025} in the hope that it will be useful for other people in this field. 
The code library used to control the devices for all measurements can be found in \cite{lonard_microscope_2025}.

\section*{Acknowledgements}
This project was funded by the Deutsche Forschungsgemeinschaft (DFG, German Research Foundation) Project-ID No. 454931666, the German Federal Ministry of Education and Research (BMBF) as part of the QuanTEAM project (FKZ13N16467), and the Quanten-Initiative Rheinland-Pfalz (QUIP).

\appendix

\section{Derivation of formulas for the magnetic-field calculation}\label{sec:b_deriv}
We derive Equations (\ref{equ:b_abs}) and (\ref{equ:b_theta}) from the characteristic polynomial, which reads
\begin{equation}
    0 = \det \begin{vmatrix}
        \lambda - \frac{1}{3}D - \mathcal{B}\cos\theta & -\frac{1}{\sqrt{2}}\mathcal{B}\sin\theta & -E                                  \\
        -\frac{1}{\sqrt{2}}\mathcal{B}\sin\theta      & \lambda + \frac{2}{3}D         & -\frac{1}{\sqrt{2}}\mathcal{B}\sin\theta      \\
        -E                                  & -\frac{1}{\sqrt{2}}\mathcal{B}\sin\theta & \lambda - \frac{1}{3}D + \mathcal{B}\cos\theta \\
    \end{vmatrix}
\end{equation}
\begin{multline}
    \Rightarrow 0 = \lambda^3 - \left(\frac{1}{3}D^2 + E^2 + \mathcal{B}^2\right)\lambda \\
    - \frac{1}{2}D\mathcal{B}^2\cos(2\theta) - \frac{1}{6}D\mathcal{B}^2 + \frac{2}{27}D^3 - E\mathcal{B}^2\sin^2\theta - \frac{2}{3}DE^2
\end{multline}
Generally, a polynomial of third order with three roots $\lambda_0, \lambda_1, \lambda_2$ can be written as
\begin{multline}
    (\lambda - \lambda_0)(\lambda - \lambda_1)(\lambda - \lambda_2) = \\
    \lambda^3 - (\lambda_0+\lambda_1+\lambda_2)\lambda^2 + (\lambda_0\lambda_1+\lambda_1\lambda_2+\lambda_0\lambda_2)\lambda - \lambda_0\lambda_1\lambda_2
\end{multline}
We define the resonance frequencies $f_\text{u}$ and $f_\text{l}$ in terms of differences between the energy levels
\begin{equation}
    \lambda_0 = \frac{1}{3}(-f_\text{u}-f_\text{l}), \quad \lambda_1 = \frac{2}{3}(f_\text{l}-f_\text{u}), \quad \lambda_2 = \frac{2}{3}(f_\text{u}-f_\text{l}),
\end{equation}
so that $f_\text{u} = \lambda_2-\lambda_0$ and $f_\text{l} = \lambda_1-\lambda_0$ and the sum of all roots vanishes $\lambda_0+\lambda_1+\lambda_2 = 0$ as expected.

The linear term of the characteristic polynomial has to fulfill 
\begin{multline}
    p = \lambda_0\lambda_1 + \lambda_1\lambda_2 + \lambda_0\lambda_2 \\
    \Rightarrow -\left(\frac{1}{3}D^2 + E^2 + \mathcal{B}^2\right) = -\frac{1}{3}(f_\text{u}^2 + f_\text{l}^2 - f_\text{u}f_\text{l}) \\
    \Rightarrow \mathcal{B}^2 = \frac{1}{3}(f_\text{u}^2 + f_\text{l}^2 - f_\text{u}f_\text{l} - D^2 - 3E^2)
\end{multline}
Additionally, the constant term of the characteristic polynomial has to fulfill 
\begin{multline}
    q = -\lambda_0\lambda_1\lambda_2 \\
    \Rightarrow -\frac{1}{2}D\mathcal{B}^2\cos(2\theta) - \frac{1}{6}D\mathcal{B}^2 + \frac{2}{27}D^3 - E\mathcal{B}^2\sin^2\theta - \frac{2}{3}DE^2 \\
    = \frac{-2f_\text{l}^3+3f_\text{l}^2f_\text{u}+3f_\text{l}f_\text{u}^2-2f_\text{u}^3}{27}.
\end{multline}
Sorting all $\theta$ to the left, dividing by $\mathcal{B}^2$, and using the expanded cosine form $\frac{1}{2}D\cos(2\theta) + E\sin^2\theta = -\frac{1}{2}D + E + D\cos^2\theta - E\cos^2\theta$ leads to 
\begin{multline}
    \Rightarrow \cos^2\theta = \frac{2f_\text{l}^3-3f_\text{l}^2f_\text{u}-3f_\text{l}f_\text{u}^2+2f_\text{u}^3}{27(D-E)\mathcal{B}^2} \\
    + \frac{2D^3 - 18DE^2}{27(D-E)\mathcal{B}^2} + \frac{D-3E}{3(D-E)}. 
\end{multline}

\begin{figure*}
    \centering
    \includegraphics[width=\linewidth]{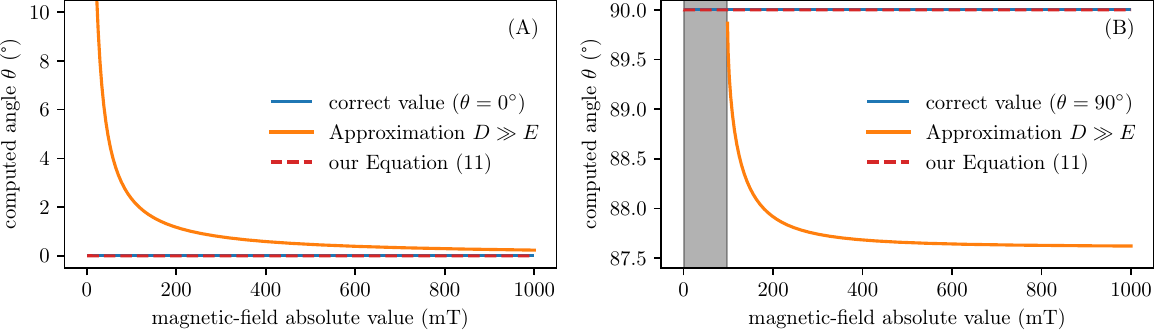}
    \caption{
        Comparison of our Equation~(\ref{equ:b_theta}) (dashed red), a possible approximation using $D \gg E$ used in \cite{balasubramanian_nanoscale_2008} (orange) (assuming $D = \SI{2870}{\mega\hertz}$ and $E = \SI{5}{\mega\hertz}$), and the expected correct value (blue). 
        In (A), the magnetic-field vector is aligned parallel to the NV axis, such that $\theta = \SI{0}{\degree}$, and in (B) the magnetic-field vector is aligned perpendicular to the NV axis, i.e., $\theta = \SI{90}{\degree}$. 
        Our Equations are, as stated, without any approximations and are therefore always exact for every magnetic-field value and angle. 
        Commonly used approximations, such as $D \gg E$, however, lead to significant systematic errors. 
        In some cases, the approximation can even lead to regions where the formulas become unsolvable, as seen in the gray area in this example.
    }
    \label{fig:angle_comparison}
\end{figure*}

The derivation of this Equation does not rely on any approximations. 
To highlight the importance of this fact, we plot in Figure~\ref{fig:angle_comparison} our Equation~(\ref{equ:b_theta}) in comparison to a formula that might be derived by using the approximation $D \gg E$. 
Suppose one, for example, starts with the same Hamiltonian as Equation~(\ref{equ:hamiltonian}), but does not rotate the magnetic-field vector into the $xz$-plane. In that case, the Hamiltonian is a function of both polar angle $\theta$ and azimuthal angle $\phi$. 
Therefore, the resulting formulas would have three unknowns (i.e. $\mathcal{B}$, $\theta$, and $\phi$) with only two known data points (i.e. $f_\text{u}$ and $f_\text{l}$). 
This set of equations is then not solvable. 
To rectify this, one might apply the approximation $D \gg E$ to eliminate the azimuthal angle $\phi$ from these equations, as seen in \cite{balasubramanian_nanoscale_2008}. 

However, as shown in Figure~\ref{fig:angle_comparison}, this approximation results in significant systematic errors. 
For small absolute magnetic-field values and small angles, the approximated formula diverges towards infinity, leading to an infinitely large error. 
For small magnetic-field values and large angles, the approximated formulas in \cite{balasubramanian_nanoscale_2008} become unsolvable, indicated by the gray background in Figure~\ref{fig:angle_comparison}.

\section{Calibration of the diamond orientation in our setup}\label{sec:field_calcs}
\begin{figure*}
    \centering
    \includegraphics[width=\linewidth]{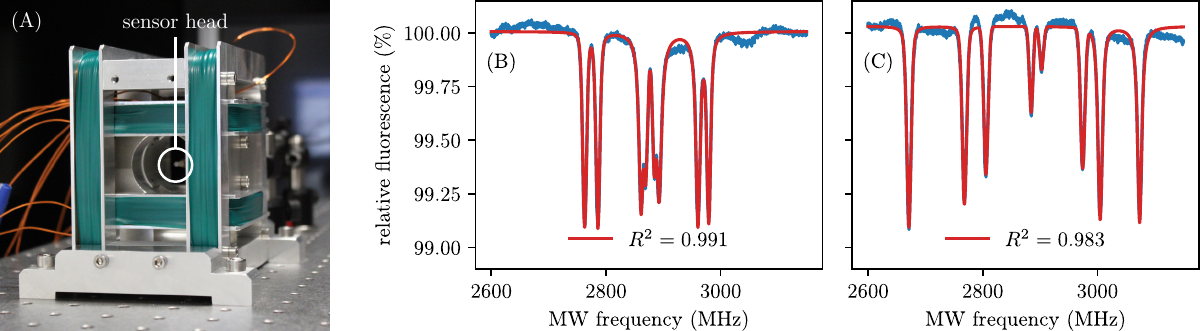}
    \caption{
        (A) The sensor previously presented in Reference \cite{dix_miniaturized_2024} was used to record the ODMR spectra shown in Figure \ref{fig:splittings}. 
        We recorded two additional spectra used for calibration when (B) only the $y$-coil is turned on with \SI{4}{\ampere} and (C) only the $z$-coil is turned on with \SI{4}{\ampere} coil current. 
        We find that the calculated magnetic-field vectors of these two spectra are not perfectly aligned at a \SI{90}{\degree} angle. 
        We reason that this alignment error of our coil system is the cause for the slight discrepancies between the theory curves and the measured resonance frequencies in Figure \ref{fig:splittings} (B). 
    }
    \label{fig:odmr}
\end{figure*}

In the experiment setup used to measure the ODMR spectra for Figure \ref{fig:splittings}, we do not know the orientation of the diamond on our sensor. 
We need one calibration measurement first to calculate the orientation of the diamond. 
For the measurement in Figure \ref{fig:splittings} (A), where the magnetic field direction is fixed and the coil current is varied, we chose the ODMR spectrum with maximal field strength as the calibration measurement. 
The theory curves are then calculated by keeping the four angles to the NV axes constant and varying the absolute value of the magnetic field. 

For the measurement in Figure \ref{fig:splittings} (B), we first recorded two ODMR spectra shown in Figure \ref{fig:odmr} with only the $z$-coil and the $y$-coil turned on, respectively. 
From these spectra, we calculate the magnetic-field vectors of the $y$- and $z$-coil using Equation (\ref{equ:bfield}) as
\begin{align}
    \mathbf{B}_{y\text{-coil}} &= 
    \SI{1.03}{\milli\tesla}\begin{pmatrix} 
        0.74 \\  
        0.66 \\  
        0.14 \\  
    \end{pmatrix} \\
    \mathbf{B}_{z\text{-coil}} &= 
    \SI{1.90}{\milli\tesla}\begin{pmatrix} 
        0.81 \\  
        0.47 \\  
        0.34 \\  
    \end{pmatrix}
\end{align}
per \SI{1}{\ampere} coil current. 
For the shown measurement, the coil currents were scaled accordingly so that the absolute magnetic-field value stays constant during the rotation of the field vector. 
Each one of these two vectors could be expanded with the $\text{O}_{\text{h}}$ symmetry, but we have the additional information that both vectors should have a \SI{90}{\degree} angle between them. 
However, when comparing all possible vector solutions, the best matching vectors have an angle of \SI{\approx 98}{\degree}. 
This discrepancy is likely due to uneven coil windings or slight alignment deviations of the two coils. 
We reason that this is the cause for the slight discrepancies between the theory curves and the measured resonance frequencies. 
Nonetheless, we rotate our coordinate system so that the $y$-coil and $z$-coil field vectors are as best aligned with the laboratory frame's $y$- and $z$-axes as possible. 
With this rotation, we then calculate the angle-dependent theory curves.

\section{Symmetry of the solutions of absolute vector magnetometry}\label{sec:symmetry}
The 48 transformation matrices of the $\text{O}_{\text{h}}$ symmetry group are derived from the six $3 \times 3$ permutation matrices with all $2^3$ possible sign combinations
\begin{align}
    &\begin{pmatrix}
        \pm 1 & 0 & 0 \\
        0 & \pm 1 & 0 \\
        0 & 0 & \pm 1 \\
    \end{pmatrix}
    , \enspace
    \begin{pmatrix}
         0 & \pm 1 & 0 \\
        0 & 0 & \pm 1 \\
        \pm 1 & 0 & 0 \\
    \end{pmatrix}
    , \enspace
    \begin{pmatrix}
        0 & 0 & \pm 1 \\
        \pm 1 & 0 & 0 \\
        0 & \pm 1 & 0 \\
    \end{pmatrix}, \\
    &\begin{pmatrix}
        0 & \pm 1 & 0 \\
        \pm 1 & 0 & 0 \\
        0 & 0 & \pm 1 \\
    \end{pmatrix}
    , \enspace
    \begin{pmatrix}
        0 & 0 & \pm 1 \\
        0 & \pm 1 & 0 \\
        \pm 1 & 0 & 0 \\
    \end{pmatrix}
    , \enspace
    \begin{pmatrix}
        \pm 1 & 0 & 0 \\
        0 & 0 & \pm 1 \\
        0 & \pm 1 & 0 \\
    \end{pmatrix}.
\end{align}

\section{Computation benchmarks}\label{sec:benchmarks}
In Figure \ref{fig:comptime}, each of the $\SI{600}{}$ data points consists of an average over $\SI{500}{}$ runs. 
Benchmarks are written in Python 3.12.0 using the \textit{timeit} package and were run on an Intel i3-12100 processor on Windows 11, version 23H2.
Enabling or disabling Python's Garbage Collector did not affect any results. 

The benchmark for the numerical approach is implemented as follows, similar to \cite{garsi_three-dimensional_2024}. 
An initial guess of a magnetic-field vector is rotated onto the four diamond-lattice axes. 
With these rotated vectors, the four Hamiltonians for each axis are computed numerically with the \textit{numpy.linalg.eigvalsh} function of the Python \textit{numpy} package \cite{noauthor_numpy_2025}. 
The SSR between the calculated and the measured resonance frequencies is then minimized by a gradient-descent algorithm that iteratively adjusts the input magnetic-field vector. 
The code for both Benchmarks can be found on Zenodo \cite{lonard_data_2025}.

\newpage


%

\end{document}